# Large decrease in the critical temperature of superconducting LaFeAsO$_{0.85}$ compounds doped with 3% atomic weight of nonmagnetic Zn impurities

Y. F. Guo,[1,2,*] Y. G. Shi,[1,2] S. Yu,[3] A. A. Belik,[1,2] Y. Matsushita,[4] M. Tanaka,[4] Y. Katsuya,[5] K. Kobayashi,[4] I. Nowik,[6] I. Felner,[6] V. P. S. Awana,[1,7] K. Yamaura,[2,3] E. Takayama-Muromachi,[1,2,3]

[1] International Center for Materials Nanoarchitectonics (MANA), National Institute for Materials Science, Tsukuba, Ibaraki 305-0044, Japan

[2] JST, Transformative Research-Project on Iron Pnictides (TRIP), 5 Sanbancho, chiyoda-ku, Tokyo 102-0075, Japan

[3] Superconducting Materials Center, National Institute for Materials Science, 1-1 Namiki, Tsukuba, 305-0044 Ibaraki, Japan

[4] NIMS Beamline Station at SPring-8, National Institute for Materials Science, 1-1-1 Kouto, Sayo-cho, Sayo-gun, Hyogo 679-5148, Japan

[5] SPring-8 Service Co. Ltd., 1-1-1 Kouto, Sayo-cho, Sayo-gun, Hyogo 679-5148, Japan

[6] Racah Institute of Physics, The Hebrew University, Jerusalem, 91904, Israel

[7] National Physical Laboratory, Dr. K.S. Krishnan Marg, New Delhi 110 012, India

We observed a large decrease of $T_c$ by no more than 3 at.% of Zn doped to the optimized superconductor LaFeAsO$_{0.85}$ ($T_c$ = 26 K), confirmed by measurements of electrical resistivity, magnetic susceptibility, specific heat, Mössbauer spectroscopy, Hall coefficient, and an electron probe micro-analysis. The rate ~9 K/% is remarkably higher than observations regarding nonmagnetic impurities. The $T_c$ suppression is likely due to pair-breaking caused by scatterings associated with highly localized electronic state of Zn doped into the Fe$_2$As$_2$ layer. If this is true, the Zn result well accords with the theoretical prediction that suggests a sign reversal $s$-wave pairing model for the Fe pnictide superconductors, unlike other nonmagnetic impurity results.

I. Introduction

   Superconductivity (SC) found in the quaternary oxyarsenide LaFeAsO$_{1-\delta}$F$_y$ resumed great activities toward discovery of a new superconducting compound in scientific communities because of prospect for achieving higher $T_c$ than that of the Cu oxides [1]. Within few years after the discovery, the $T_c$ went over 50 K by replacing La to other rare-earth element, and materials varieties were rapidly developed: many Fe pnictides containing the Fe$_2$As$_2$ layer in each structure have been proved to become superconducting, e.g., $A$Fe$_2$As$_2$ ($A$ = alkaline earth) [2], $A$FeAsF [3], and Sr$_4$Sc$_2$Fe$_2$P$_2$O$_6$ [4]. It is noteworthy that the Fe-based superconductors may have practical values because their upper-critical field is remarkably high [5].

   The SC in the Fe pnictides emerges after the spin-density wave (SDW) is suppressed by chemical or physical methods, being analogical with the SC of the Cu oxides [6]. The Fe$_2$As$_2$ layer is believed to play a decisive role in establishing the SC as does the CuO$_2$ layer [7]. These facts imply the Fe pnictides and the Cu oxides may share common physics in part regarding mechanism of the SC. To date tremendous activities were performed to reveal nature of the SC of the Fe pnictides; however, the nature seems not to be well uncovered yet. For example, the SC gap symmetry is still under debate. In early theoretical studies, there seems to be a consensus that the gap symmetry is of a sign-reversal s-wave (often notated as $s_\pm$-wave) as proposed independently by Mazin *et al.* [8], Kuroki *et al.* [9,10], and Tesanovic *et al.* [11,12]. Probable nesting between separated pockets of the Fermi surfaces may enhance spin fluctuations, helping to generate the SC. In the early studies, the SC gap has no nodes; full gap opening was predicted [13,14]. The gap symmetry model was further supported by experiments including NMR studies [13,15,16], angle-resolved photoemission spectroscopy [17,18], microwave penetration depth measurements [19], μSR studies [20,21], and neutron scattering studies [22,23] on varieties of the Fe-based superconductors.

   However, the pairing model seems not to meet results of impurity studies: many doped elements to the Fe-based superconductor rather work weakly for the SC. For example, Co, Ni, Ru, Rh, Pd, and Ir were found to even work to produce the SC, suggesting that the Fe-based superconductors are highly robust against impurities [24-27]. The observations seemed to sharply contrast to what is expected for the $s_\pm$-wave SC [28,29], since the $s_\pm$-wave SC is supposed to be quite fragile against even a nonmagnetic impurity unlike the conventional *s*-wave SC (BCS SC is normally robust against nonmagnetic impurities as observed for MgB$_2$ for instance [30]).

   Additional theoretical studies, however, suggested that the $s_\pm$-wave model can be possibly consistent with the robust SC if certain conditions such as a small impurity potential (<<1 eV) and a large potential radius are effective in the SC [28,29]. Nevertheless, the issue regarding inconsistency between the $s_{\pm}$-wave model and the results of the impurity studies seems to eventually complicate discussions about the mechanism of the SC of the Fe pnictides, and the issue is thus subjected for further studies. To our best knowledge, a fragile SC against a nonmagnetic impurity was not observed yet for any Fe pnictide superconductors at the time of submission of this article [31,32]. Before the high-pressure synthesis of $LaFeAsO_{0.85}$ [33], it appeared that the doped Zn to $LaFeAsO_{1-x}F_x$ does not reduce $T_c$ (< 10 at. % Zn) [31]. Meanwhile, a study on the magnetic Mn doped $LaFeAsO_{1-x}F_x$ found a large $T_c$ decrese, while the nonmagnetic Co doped $LaFeAsO_{1-x}F_x$ slowly decreses $T_c$ over the doping [32,34].

In this paper, we report a large $T_c$ decrease achieved by a minimal level of Zn (below 3 atomic %) doped into the superconductor $LaFeAsO_{0.85}$ [33]. We discuss role of Zn in the SC, and compare the result with other results reported elsewhere.

## II. Experimental

Polycrystalline samples of Zn-doped $LaFe_{1-x}Zn_xAsO_{0.85}$ ($x$ = 0, 0.005, 0.01, 0.015, 0.02, 0.03, 0.05) were prepared by solid-state reaction under high pressure from powders of LaAs (lab-made), $Fe_2O_3$ (3N, Furuuchi Chem. Co.), ZnO (3N, Wako), and Fe (3N, 100 mesh, Rare Metallic Co.). LaAs was prepared in advance from La pieces (3N, Nilaco Co.) and As powder (5N, High Purity Chem.) with 1 at.% excess As by heating in an evacuated quartz tube at 500 ºC for 20 h, followed by quenching to room temperature. The LaAs product was ground and re-heated at 850 ºC for 10 h. The lab-made LaAs powder was qualitatively studied by a method of powder x-ray diffraction (XRD) using Cu-K$\alpha$ radiation in a diffractometer in RINT2200V/PC, Rigaku.

The starting mixture each was placed into an h-BN cell (preheated in advance at ~2000 ºC for 1 h in nitrogen) and the cell was sealed in a gold capsule. The sample-contained capsule was each heated at 1300 ºC for 2 h in a belt-type pressure apparatus, which is capable of maintaining 6 GPa during the heating, followed by quenching to room temperature before releasing the pressure. To increase uniformity of the sample, the obtained pellet was carefully ground and re-heated at the same condition.

The samples of the oxygen vacancy-controlled $LaFeAsO_{1-\delta}$ ($\delta$ = 0, 0.12, 0.15, 0.22) were prepared in the same way under high pressure, except the $\delta$ = 0 sample. The $\delta$ = 0 sample was synthesized in an evacuated quartz tube at 1100 ºC for 30 h under ambient pressure because the pressurized conditions did not work better to improve the sample quality. The poor quality persisted exceptionally at $\delta$ = 0 (not due to a run-to-run error) as observed in the high pressure synthesis of TbFeAsO$_{1-\delta}$ [35]. It should be emphasized that the optimally carrier doped samples ($\delta$ = 0.15) with and without Zn were all prepared under the same high-pressure condition.

All samples were qualitatively studied by the XRD method, and selected samples (LaFeAsO$_{0.85}$ and LaFe$_{0.99}$Zn$_{0.01}$AsO$_{0.85}$) were further investigated by a synchrotron X-ray diffraction (SXRD) method. The SXRD measurement was conducted at $\lambda$ = 0.652973 Å in a large Debye-Scherrer camera at the BL15XU beam line of SPring-8 [36]. The sample capillary, Lindenmann glass, was rotated during the measurement. The Rietveld analysis was carried out by using RIETAN-2000 [37]. The samples of LaFeAsO$_{0.85}$ and LaFe$_{0.99}$Zn$_{0.01}$AsO$_{0.85}$ were further studied by a Mössbauer spectroscopy, which was carried out at room temperature by using a conventional constant acceleration drive and a 50 mCi $^{57}$Co:Rh source. The experimental spectra were analyzed by a least-squares fit procedure. The velocity calibration and isomer shift (IS) zero are those of $\alpha$-Fe measured at room temperature.

A selected Zn doped sample (LaFe$_{0.95}$Zn$_{0.05}$AsO$_{0.85}$) was studied in an electron probe micro-analysis (EPMA) at an acceleration voltage of 15 kV (JXA-8500F, JEOL). A surface of the sample pellet was carefully polished using an alumina coated sheet. The analysis confirmed that possible contaminations from such as Au are below the EPMA background level. Expected peaks due to Zn were confirmed in a wavelength-dispersive spectrometry, indicating Zn is indeed incorporated in the sample. The same sample was further studied by an element mapping operation in EPMA regarding Zn over a square surface approximately 100 μm × 100 μm maximum in area size. The doped Zn was observed to evenly spread out, suggesting that Zn-rich impurities are unlikely formed. Additional EPMA operations were carefully conducted regarding possible formation of the Zn-rich impurities; however any trace of the possibility was undetected.

The magnetic susceptibility ($\chi$) of the samples was measured in the magnetic property measurement system, Quantum Design Inc. Loose powder was cooled to 2 K before applying a magnetic field (zero-field cooling; ZFC), followed by warming to 300 K in a magnetic field 10 Oe. The sample was then cooled down to 2 K in the field (field cooling; FC). The electrical resistivity ($\rho$) was measured in the physical properties measurement system (PPMS), Quantum Design Inc., by a four-probe method with a constant gauge current of 0.2 mA. Hall coefficient ($R_H$) was measured by rotating the sample by 180 degrees in a field of 50 kOe in PPMS between 25 K and 300 K. Specific heat ($C_p$) of the samples LaFe$_{1-x}$Zn$_x$AsO$_{0.85}$ ($x$ = 0 and 0.05) was measured in PPMS, between 2.2 K and 300 K by a heat pulse relaxation method.

**III. Results and discussion**

The powder XRD patterns at $\delta = 0.15$ (LaFe$_{1-x}$Zn$_x$AsO$_{0.85}$) are shown in Fig. 1a. All peaks were well indexed by assuming the ZrCuSiAs-type structure with *P*4/*nmm* as was done for TbFeAsO$_{0.85}$ [38]. Although the XRD analysis indicated that the samples are of high-quality, the SXRD analysis (Figs. 1b and 1c) found a tiny amount of LaAs incorporated in the sample regardless of amount of Zn. The incorporated LaAs was observed in the EPMA. The impurity may result from in part the compositional issue of the starting powder of LaAs prepared with excess As.

Rietveld analysis of the SXRD patterns was carefully carried out and the results are shown in Figs. 1b and 1c. A reliable structure solution was obtained [39]. The mean La-As distance is 3.3577(6) at $x = 0$ and 3.3570(6) at $x = 0.01$, and the Fe-As-Fe angle in the Fe$_2$As$_2$ layer is 113.09(9)° and 113.20(9)°, respectively. Although La-As distance and the angle were suggested to play a crucial role in controlling the effective bandwidth, thereby affecting the SC [40], those were confirmed to change quite little over the Zn substitution.

The lattice parameters of the tetragonal unit cell, deduced from the XRD patterns, were plotted against the Zn concentration as shown in Figs. 2a-2c. For a comparison, the lattice parameters of the oxygen vacancy controlled samples LaFeAsO$_{1-\delta}$ were plotted along the Zn doped data. Regarding LaFeAsO$_{1-\delta}$, the tetragonal lattice parameters *a* and *c* decrease with increasing $\delta$ as well as what was found for the compounds TbFeAsO$_{1-\delta}$ [35]. This probably reflects combination of Coulomb attractive forces between the charged [LaO$_{1-\delta}$]$^{1+2\delta}$ and [FeAs]$^{1-2\delta}$ layers and amount of vacate sites for oxygen atoms. In sharp contrast to the observation, the Zn substitution resulted in an anisotropic change of the lattice parameters: *c* of LaFe$_{1-x}$Zn$_x$AsO$_{0.85}$ increases (+0.02 % at Zn$_{0.01}$), while *a* decreases much efficiently (-0.08 %) over the substitution. The distinctive feature is indicative of influence of the doped Zn over the unit cell. Besides, magnetic $T_c$ vs. *c/a* (Fig. 2d) indicates a tight relation between those (magnetic data are shown later). These facts clearly indicate that the Zn substitution is successful up to 5 at.% under the high-pressure condition. The EPMA analysis entirely supports this in fact. Usually, loss of ZnO becomes significant during heating above 1100 °C by volatility [41], thus experimental chances of Zn substitution to an oxide is often limited. Perhaps, the high-pressure condition might allow us to reduce the ZnO volatility during heating, resulting in the successful Zn substitution to LaFeAsO$_{0.85}$.

Additionally, selected Zn doped samples were further characterized by measurements of Hall coefficient, conducted from 300 K to 25 K. The data are shown in Fig. 3 for the samples with and without 2 at.% doped Zn. The features are essentially identical, indicating that the electron count changes little over the Zn substitution. The carrier concentration at 300 K for LaFeAsO$_{0.85}$ is $3.9 \times 10^{21}$ cm$^{-3}$, corresponds to 1.1 electrons per the primitive cell, being comparable with the values reported for the F-doped LaFeAsO [42].

Let us see the superconducting properties of the Zn doped samples of LaFe$_{1-x}$Fe$_x$AsO$_{0.85}$. Figs. 4a-4g show Zn concentration dependence of $\chi$ vs. $T$. In Fig. 4a, the Zn-free sample with optimally carrier doped ($\delta$ = 0.15) clearly undergoes a superconducting transition at ~26 K as was reported elsewhere [33]. Employing the calculated density of 7.85 g/cm$^3$, the magnetic shielding fraction is estimated to be 1.13 (1.00 is expected for the perfect shielding), indicating homogeneous SC of the sample. In contrast, the Meissner fraction (5 K, FC curve) is fairly small, less than 0.1. The contrastive features between the magnetic shielding and the Meissner fractions were commonly observed for the oxygen-deficient 1111 systems [38] probably because of possible efficient magnetic-flux pins and small particle size not far from the magnetic penetration depth (~0.25 µm [43]). In fact, we observed that the particles are much smaller than 1 µm in the EPMA.

Increasing the Zn concentration, $T_c$ quickly goes down; no more than 3 at.% of Zn almost completely suppresses the SC. The $T_c$ variation is summarized in Fig. 4h. The feature highly contrasts with what was observed for the Co doped LeFeAsO$_{1-x}$F$_x$ [32,34] and is nearly comparable with the magnetic Mn doped LeFeAsO$_{1-x}$F$_x$ [32]. Besides, the nonmagnetic Zn result is highly contrastive with what were observed by $d$ elements doping studies such as Ni, Ru, Rh, Pd, and Ir, which weakly suppresses $T_c$ or even produces the SC [24-27]. Moreover, the large $T_c$ decrease by Zn is not at all comparable with the recent result for the Zn doped LeFeAsO$_{1-x}$F$_x$ [31], thus we tested a possible run-to-run error in our synthesis. We actually repeated the synthesis of all the LaFe$_{1-x}$Fe$_x$AsO$_{0.85}$ samples (except $x$ = 0.05) 4 times in total. An independent set of the magnetic data are for instance shown in Figs. 4a-4g, clearly confirming reproducibility of the essential part of the result.

Figs. 5a and 5b show the temperature dependence of $\rho$ of LaFe$_{1-x}$Zn$_x$AsO$_{0.85}$ and LaFeAsO$_{1-\delta}$, respectively. Regarding the stoichiometric LaFeAsO, $\rho$ at 300 K is ~6 mΩcm, being comparable with the normalized mean free path $k_F l \approx 0.6$ [44]. The $\rho$ of LaFeAsO gently varies with a broad minimum at approximately 220 K and pronouncedly goes down at 150 K, corresponding to the SDW instability [44]. By introducing the oxygen vacancies, the normal-state $\rho$ becomes much smaller, reflecting increase of the carrier density in the Fe$_2$As$_2$ layer. The residual resistivity ratio [RRR; $\rho(300\ K)/\rho(T_c)$] at the optimized carrier density for SC (LaFeAsO$_{0.85}$) is ~6, being not far from the values reported for the Co doped LaFeAsO [34].

$T_c$ determined by the $\rho$ measurement steeply goes to zero by the Zn substitution (Fig. 5a) in well accord with the magnetic susceptibility measurements. Regarding the normal stare $\rho$, the quadratic like temperature dependence and the prominent upturn on cooling were observed in the Zn-doped samples (see for example at $x = 0.01$) and RRR is deteriorated somewhat up to $x$ of 0.01, probably reflecting enhanced scattering factors. The further doped $x = 0.02$ sample, however, shows anti-direction features, while the SC remains to be suppressed. It is possible that the doped Zn works not only as a scattering center for charges but also to weaken magnetism of the $Fe_2As_2$ layer, resulting in the slight improvement of the charge transport as a result of delicate balance of those factors. The $x = 0.03$ and 0.05 $\rho$ curves show a weak charge localization, which is probably in the Anderson's scheme [45], as the $\rho$ remains low and the temperature dependence is weak. The increased amount of Zn in the conducting $Fe_2As_2$ layer may further develop degree of the scatterings to be responsible for the weak localization. It should be noted that the sharp drop of $\rho$ at $x = 0.03$ is not due to a bulk SC because corresponding magnetic transition is absent in the $\chi$ vs. $T$ measurements. Additional studies on high-quality single crystals if available would be helpful to further address role of Zn.

We considered a possibility that the drastic suppression of $T_c$ in the Zn-doped samples is due to the carrier localization rather than the pair breaking. Thus, we checked the present data on the point. We found that the normal state $\rho$ of the $T_c$ suppressed $LaFe_{1-x}Zn_xAsO_{0.85}$ ($x = 0.02$) (Fig. 5a) is much smaller than that of the superconducting sample $LaFeAsO_{1-\delta}$ ($\delta = 0.12$, $T_c = 21$ K) (Fig. 5b), even though both the data were taken from the sintered polycrystalline forms. The entire systematic change of the normal state $\rho$ against compositions $\delta$ and $x$ reasonably suggests that the difference is substantial beyond the polycrystalline nature. The comparison suggests that the electron localization picture is unlikely for the large $T_c$ decrease.

Fig. 6 shows $C_p/T$ vs. $T$ plots for the optimized SC samples ($LaFeAsO_{0.85}$) and the Zn doped non-SC sample ($LaFe_{0.95}Zn_{0.05}AsO_{0.85}$). It appears that an expected anomaly at $T_c$ is unfortunately unclear in the plot. It is possible that disorder regarding oxygen vacancy distribution causes inhomogeneous SC states, much broadening the expected peak, and the broad anomaly is masked by the lattice contributions. For instance, remarkably broad anomaly was observed for the Co doped LaFeAsO [34] and the oxygen vacant $LaFeAsO_{1-\delta}$ [35]. The F-doped LaFeAsO having comparable $T_c$ with $LaFe_{0.95}Zn_{0.05}AsO_{0.85}$ actually does not show even a broad anomaly at $T_c$ [46].

We analyzed the low-temperature part of $C_p$ (<15 K) of the samples with and without doped Zn. Using the approximate Debye model $C(T)/T = \beta T^2 + \gamma$, where $\beta$ is a coefficient and $\gamma$ is the Sommerfeld coefficient, we obtained $\beta$ of $2.79(1) \times 10^{-4}$ J mol$^{-1}$ K$^{-4}$ and $\gamma$ of 2.0(1) mJ mol$^{-1}$ K$^{-2}$ for $LaFeAsO_{0.85}$ by a fit to the linear part (inst to Fig. 6). We obtained $T_D$ (the Debye temperature) of 299(1) K from $\beta = 12\pi^4 R/5T_D^3$. The $\gamma$ and $T_D$ are indeed comparable with those of the Co- and F-doped LaFeAsO superconductors [34,46]. It is most noteworthy that the Zn doped sample shows additional anomaly at the low temperature limit; a broad upturn appears on cooling and $C_p/T$ approaches to ~12 mJ mol$^{-1}$ K$^{-2}$. Although we expected a Schottky anomaly in the data because of possible Zn-induced magnetic moments in the Fe$_2$As$_2$ layer, the broad upturn was eventually observed instead. We found that a comparable upturn was observed for the Zn doped YBCO, which was discussed using Kondo-screened moments [47]. Our observation for the low temperature $C_p$ of LaFe$_{0.95}$Zn$_{0.05}$AsO$_{0.85}$ is not well understood.

We attempted to measure the oxygen content of the samples LaFe$_{1-x}$Zn$_x$AsO$_{0.85}$ ($x \leq 0.02$) by a gravimetric method. About 10 mg of each sample was fully oxidized at 1500 °C in air for 24 h, followed by cooling slowly to room temperature. The final product was identified to be LaFeO$_3$ in the XRD study, indicating that the following reaction was proceeded: LaFeAsO$_{1-\delta}$ + [4.5-(1-$\delta$)]/2O$_2$ --> LaFeO$_3$ + 0.5As$_2$O$_3$↑. The net oxygen content was calculated by monitoring the weight loss. Assumed that Zn is fully evaporated in the heating, the net oxygen content is within the range 0.83 - 0.85 per the formula unit, being in good agreement with the nominal. Although uncertainty of the exact Zn amount results in a possible error, it should be smaller than 0.014 per the formula unit. It thus appears that the oxygen content variation unlikely accounts for the large $T_c$ decrease [33,35].

Additionally, we studied the Mössbauer effect of the $x = 0$ ($T_c = 26$ K) and 0.01 ($T_c = 10$ K) samples. As shown in Figs. 1b and 1c, both the spectra are nearly identical, suggesting the Fe valence is unaltered between the two samples. In more details, the hyperfine parameters and the IS values are the same and the quadrupole splitting (QSP) is negligibly small. The line width = 0.265(5) mm/s, IS(relative to Fe) = 0.451(2) mm/s, QSP = 0.090(3) mm/s for the $x = 0$ sample, and 0.277(5), 0.455(2), and 0.096(3), respectively, for the $x = 0.01$ sample. The oxygen content variation is again confirmed little.

Let us focus on the role of the doped Zn. The divalent Zn has the 3$d^{10}$ configuration; the $d$ orbital is fully occupied in contrast to Co$^{2+}$ (3$d^7$) and Ni$^{2+}$ (3$d^8$), resulting in a highly localized nature. Indeed, the Zn 3$d$ states are located at -8 to -6.5 eV far below the Fermi level in LaZnAsO [48]. It is therefore reasonable that doped Zn in the Fe site holds strong localization nature and thus does not add itinerant electrons into the Fe$_2$As$_2$ layer. It should be noted that the small $T_c$ suppression of the Co doped LaFeAsO$_{1-\delta}$F$_\delta$ may be due to weaker localization of Co [32]. Since the $T_c$ suppression is achieved by no more than 3 at.% of Zn, it is most likely that the doped Zn works as a scattering center as predicted in a recent theoretical work: the scattering may effects pair-breaking of the $s_\pm$-wave SC [49]. Our observations probably accords with the prediction.

Independent theoretical studies on a 5 $d$-orbital model also predicted that the Anderson's theorem is violated for the $s_\pm$-wave state due to strong inter-band impurity scattering, suggesting a significant nonmagnetic impurity effect on the SC [28,29]. This may explain our results for the large $T_c$ decrease achieved by the few at. % Zn.

Although our results well match with the theoretical predictions, those firmly contradict the results for the Zn doped studies on LaFeAsO$_{1-\delta}$F$_{0.1}$ [31], which concluded that the doped Zn does not suppress the SC. It was stated in Ref. 31 that the SC remains almost unperturbed or even enhanced by the Zn substitution (<10 at.%), being strictly contrasting with the present results. Afterward, the same research group, however, reported additional results (after the present article was submitted) [50], showing clearly a large $T_c$ decrease as well as what we observed for the LaFe$_{1-x}$Zn$_x$AsO$_{0.85}$. It appeared that degree of the $T_c$ suppression may depend on F concentration as discussed in Ref. 51. This may be relevant to what was found in recent NMR studies on P-doped BaFe$_2$As$_2$ [51] and theoretical studies on local structures of the Fe$_2$As$_2$ layer [10], those suggested the gap symmetry can possibly change depending on minute factors. Eventually, the fragile SC of the doped LaFeAsO was confirmed by 2 independent experiments using the nonmagnetic Zn. Thus, it is hardly solid that the 1111 superconducting compound is sensitive to the nonmagnetic Zn impurity. In contrast, the Zn doped 122 compound Ba$_{0.5}$K$_{0.5}$Fe$_2$As$_2$ was found to be insensitive to the doped Zn [52]; the issue, how the distinction between the sensitivity to the 1111 system and the insensitivity to the 122 system comes from, is left for future studies.

In summary, a large $T_c$ decrease from the optimum $T_c$ of 26 K was observed by a minimal amount of Zn (< 3 at. %) doped into LaFeAsO$_{0.85}$. The $T_c$ suppression rate of Zn was found to be ~9 K/%, being remarkably higher than the rate (~2.5 K/%) for the nonmagnetic impurity study on LaFe$_{1-y}$Co$_y$AsO$_{1-x}$F$_x$ [32] and almost comparable with the magnetic impurity study on LaFe$_{1-y}$Mn$_y$AsO$_{1-x}$F$_x$ [32]. The present data indicated that the doped Zn likely plays a dominant role of effecting pair-breaking due to scatterings associated with the highly localized state of Zn in the Fe$_2$As$_2$ layer. If this is true, the $T_c$ suppression is consistent with the prediction from the $s_\pm$-wave model. For further clarification of the role of Zn, NMR measurements are in progress. Although the present results indicate that the conventional $s$-wave model is highly unlike for LaFeAsO$_{0.85}$, the $s_\pm$-wave model and the nodal $d$-wave model both remain possible. While many theoretical studies suggested that the $s_\pm$-wave model is much likely for the Fe pnictides, the present nonmagnetic impurity study is however unable to judge which is mostly like.

**Acknowledgments**

We thank Dr. D. J. Singh for valuable discussion. This research was supported in part by the World Premier International Research Center (WPI) Initiative on Materials Nanoarchitectonics from MEXT, Japan; the Grants-in-Aid for Scientific Research (20360012, 22246083) from JSPS, Japan; and the Funding Program for World-Leading Innovative R&D on Science and Technology (FIRST Program) from JSPS.

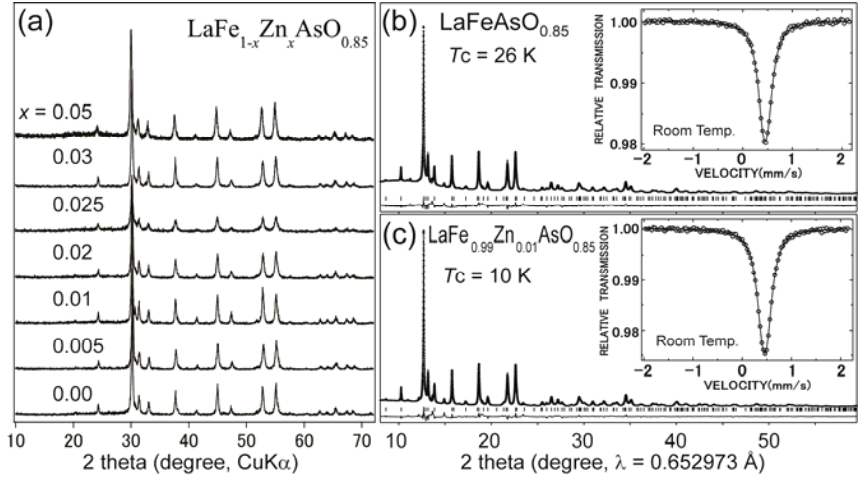

Fig. 1  Powder XRD patterns of (a) LaFe$_{1-x}$Zn$_x$AsO$_{0.85}$ ($x$ = 0 - 0.05) and Rietveld analysis of the SXRD patterns of (b) LaFeAO$_{0.85}$ and (c) LaFe$_{0.99}$Zn$_{0.01}$AsO$_{0.85}$.  Dots and lines represent the observed and the calculated intensities, respectively.  Each difference curve is shown at the bottom.  Small vertical bars indicate calculated Bragg reflection positions.  Inset each shows the Mössbauer spectrum at room temperature.

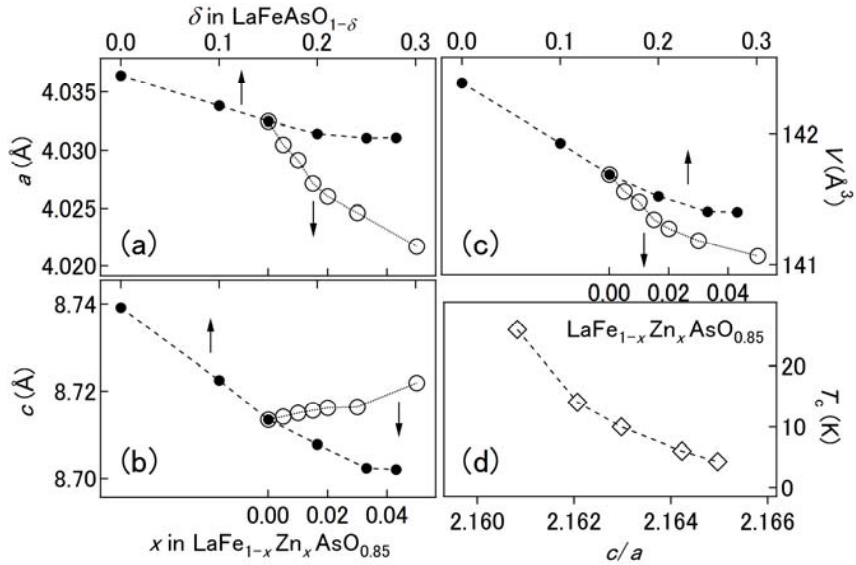

Fig. 2  (a-c) Unit-cell evolution of LaFe$_{1-x}$Zn$_x$AsO$_{0.85}$ (bottom axes) and LaFeAsO$_{1-\delta}$ (top axes, taken from ref. 33).  (d) $T_c$ vs. $c/a$.  The dashed lines are guides to the eye.

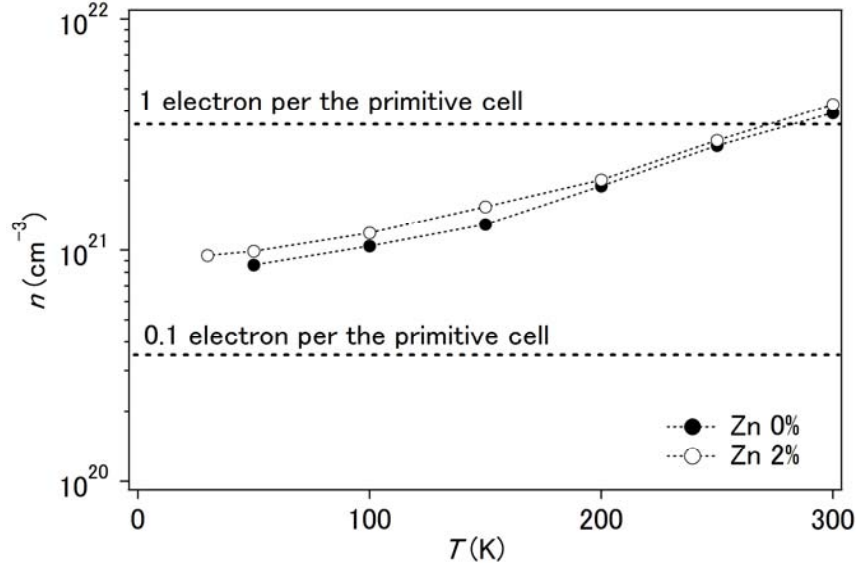

Fig. 3   $T$ dependence of the Hall number ($n = |e/R_H|$) of LaFe$_{1-x}$Zn$_x$AsO$_{0.85}$ ($x = 0$ and 0.02).   The sign of $R_H$ is negative over the $T$ range studied.

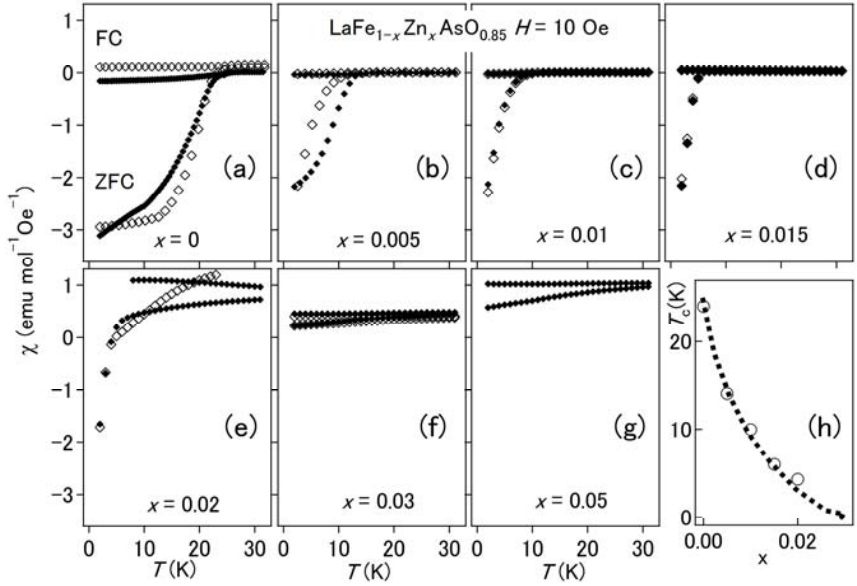

Fig. 4   (a-g) $T$ and $x$ dependence of $\chi$ of LaFe$_{1-x}$Zn$_x$AsO$_{0.85}$ measured at 10 Oe in the ZFC and FC conditions.   Open and closed symbols represent data of independent set of samples.   (h) $T_c$ vs. $x$.

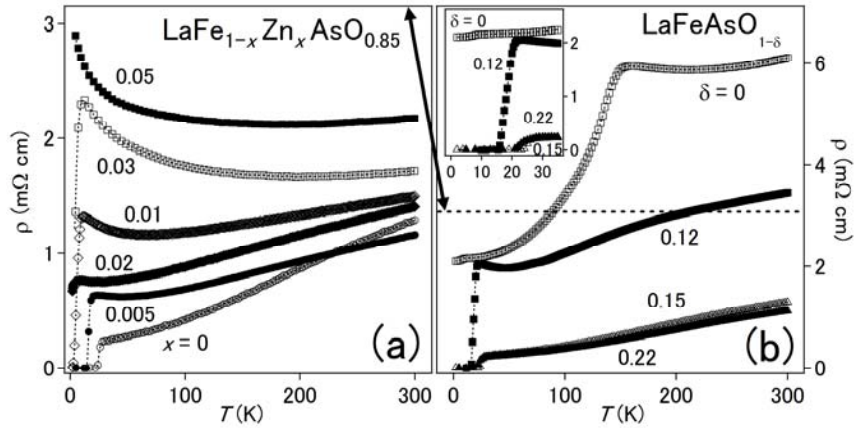

Fig. 5  $T$ dependence of $\rho$ of (a) LaFe$_{1-x}$Zn$_x$AsO$_{0.85}$ and (b) LaFeAsO$_{1-\delta}$.  Inset shows an expansion.  The dotted line corresponds to the scale indicated by the solid arrow.

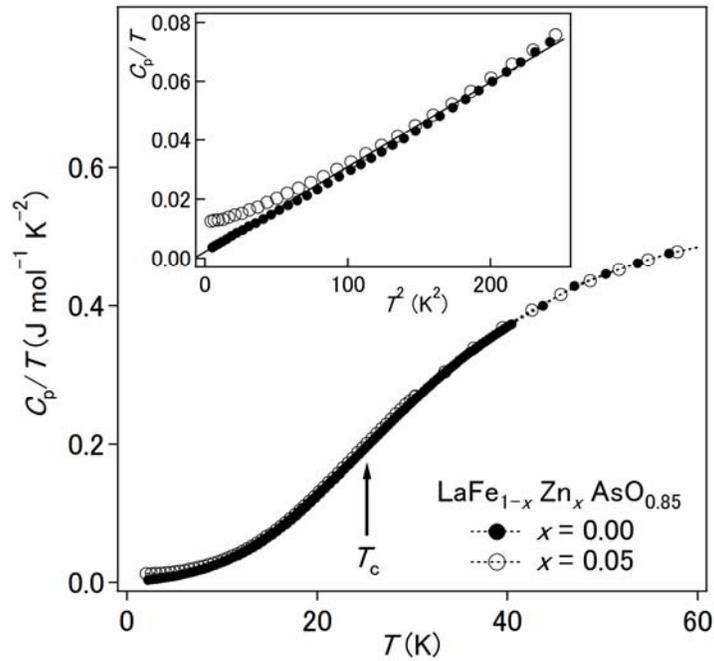

Fig. 6  $T$ dependence of $C_p$ of LaFe$_{1-x}$Zn$_x$AsO$_{0.85}$ at $x = 0.00$ and $0.05$.  Inset shows an alternative plot.  The solid line indicates a fit to the data.